\documentclass[microtype]{gtpart}
\pdfoutput=1
\usepackage{pinlabel}
\title{A geometric alternative to dark matter}

\author{Colin Rourke}
\address{Mathematics Institute\\University of Warwick\\Coventry CV4 7AL\\UK}
\email{cpr@msp.warwick.ac.uk}
\urladdr{http://msp.warwick.ac.uk/~cpr}

\subject{primary}{msc2000}{85A05}
\subject{secondary}{msc2000}{85A15}
\subject{secondary}{msc2000}{83C57}
\subject{secondary}{msc2000}{83F05}

\def\sh#1{\subsection*{#1}\addcontentsline{toc}{subsection}{#1}}
\def\astar{${\rm A}^*$}
\def\id{inertial drag}

\def\rdot{\kern 2pt\dot {\kern -1 pt r}}
\def\rdd{\text{\em\"r}}
\def\thetadot{\dot\theta}

\makeop{inert}
\def\lfrac#1#2{#1/#2}
\def\strutt{\vrule width 0pt height 20pt}
\newtheorem*{theorem}{Theorem}
\def\thinm{\mkern 2mu}
\def\id{inertial drag}
\def\wrt{with respect to}
\begin{document}
\begin{abstract}
The existence of ``dark matter'', inferred from the observed rotation
curves of galaxies, is a hypothesis which is widely regarded as
problematic.  This paper proposes an alternative hypothesis based on
the space-time geometry near a rotating body and formulated in terms
of the dragging of inertial frames.  This hypothesis is true in a
certain linear approximation to General Relativity (Sciama
\cite{Sciama}) and is justified in general by Mach's principle.  Dark
matter corrects the rotation curve but does not predict the ubiquitous
spiral structure of galaxies.  The geometric alternative suggested
here deals with both problems and allows the construction of a simple
model for the dynamics of spiral galaxies which fits observations
well.
\end{abstract}

\maketitle

\section{Introduction}\label{sec:intro} 

One of the major problems with the current consensus model for the
universe is the problem of dark matter which has no observations to
support its existence and no agreement as to its nature.  The purpose
of this paper is to propose an alternative based on the geometry of
space-time, which, like dark matter, explains observed rotation
curves.  There is a dual problem to the dark matter problem, namely
the spiral structure problem.  In the current consensus model for the
universe, with dark matter, there is no satisfactory model for
galactic dynamics that explains the ubiquitous spiral form of many
galaxies.  By contrast the geometric postulate made here fits observed
rotation curves and also provides a good model for galactic dynamics
fitting observed spiral structure.

Dragging of inertial frames, henceforth abbreviated to ``inertial
drag'' or ID, is a relativistic effect whereby the non-inertial motion
of a body (acceleration and$/$or rotation) causes the inertial frames at
other points to be dragged.  In his thesis \cite{Sciama} Sciama
proposed that this effect should embody Mach's principle (that
inertial frames depend on the total distribution of matter in the
universe) and that this should specify the full dynamics of the
universe.  He illustrated this with a special case (a certain linear
approximation to general relativity) where it all works perfectly.
His formula for the ID effect of a rotating body is this:

\rk{Weak Sciama Principle (WSP)}{\sl A mass $M$ at distance
  $r$ from a point $P$, rotating with angular velocity $\omega$,
  contributes a rotation of $k\thinm M\thinm \omega/r$ to the inertial
  frame at $P$ where $k$ is constant.}
\medskip

It is called the ``Sciama principle'' to distinguish it from the more
general ``Mach principle'' and ``weak'' because it specifies only the
effect of one rotating body, instead of all non-inertial motions.  The
``Full Sciama Principle'' (for rotation) states that the rotation of a
local inertial frame is obtained by adding the effects for all
rotating bodies in the accessible universe (ie not regressing faster
than $c$).
%\enlargethispage{20pt}(footnote rev1.2 removed)

%%%%% WEIGHTING  Rev1.3
The factor $k\thinm M/r$ must be regarded as a weighting factor and
the sum must be divided by the sum of the weights.  For an example see
the derivation of \eqref{eq:nett} below.

This principle (the WSP) is the proposed geometric hypothesis.  The
key factor for the rotation curve is $1/r$ which has several pieces of
evidence in its favour.  It fits Sciama's approximation.  It fits the
behaviour of apparent acceleration (see Mach \cite[Ch II.VI.7 (page
  286)]{Mach}).  There is a simple dimensional argument that supports
it (see the discussion of the precession of the Foucault pendulum in
Misner, Thorne and Wheeler \cite{MTW} starting on page 547 para 3 with
the margin note \emph{The dragging of the inertial frame}).

The hypothesis is obviously related to Mach's principle though it is
much weaker.  But it has the advantage that it is capable of direct
verification by observation.  Although true in an approximation to
general relativity (GR), it is not true in general in GR: in the Kerr
metric, ID drops off with $1/r^3$ not $1/r$.  Therefore to accomodate
it within GR it is necessary to assume that a rotating body has an
effect on the local field that causes the hypothesised ID effect and
stops it being a vacuum.  This effect embeds Mach's principle (for
rotation) within GR and avoids the paradoxes that arise in a naive
formulation of the principle: the ID field is a form of gravitational
disturbance that propagates, like all gravitational disturbances, at
the speed of light.

\section{The rotation curve}\label{sec:rot}

The basic assumption is that the centre of every galaxy contains a
heavy rotating mass (presumably a black hole).  It is the ID effects
from this mass that cause equatorial orbits to exhibit the
characteristic rotation curve.  However, the analysis applies to any
axially-symmetric rotating body, which does not need to be assumed to
be heavy.

To fix notation, consider a central mass $M$ at the origin in
$3$--space which is rotating in the right-hand sense about the
$z$--axis (ie counter-clockwise when viewed from above) with angular
velocity $\omega_0$.  Assume a flat background space-time, away from
$M$, with sufficient fixed masses at large distances to establish a
non-rotating inertial frame near the origin, if the effect of $M$ is
ignored.  Let $P$ be a point in the equatorial plane (the
$(x,y)$--plane) at distance $r$ from the origin.  The rotation of the
inertial frame at $P$ is given by adding the contribution from $M$ to
the contribution from the distant masses.  The inertial frame at $P$
is rotating coherently with the rotation of $M$ by the average of
$\omega_0$ weighted $kM/r$ and zero (for the distant fixed masses)
weighted $C$ say.  Normalise the weighting so that $C=1$
(which is the same as replacing $k/C$ by $k$) which leaves just one
constant $k$ to be determined by experiment or theory.  The nett
effect is a rotation of
\begin{equation}\label{eq:nett}
\frac{(kM/r)\times\omega_0 +1\times 0}{(kM/r) + 1} =
\frac{A}{r+K}\ \ \text{where}\ \ K=kM\ \ \text{and} \ \ A=K\omega_0.
\end{equation}

\rk{Note}If the full Sciama principle is assumed then $k=1$,
$K=M$ and $A=M\omega_0$.  However the choice of $k=1$ is not relevant
to the arguments presented in this paper.  Nothing that is proved
depends on knowing the exact relationship between $K$ and $M$.
When constructing models for galaxies in \fullref{sec:dyn} the value
$k=1$ will be used for definiteness and to the extent that these
models fit observations of real galaxies this provides evidence for
the full Sciama principle.
\medskip

The key to the rotation curve is to understand the way in which the
\id\ field affects the dynamics of particles moving near the origin.
For simplicity work in the equatorial plane and in a linear
approximation to a background Minkowski space.  Further assume a
Newtonian formula for the central attractive force.  For more
accuracy, an approximation to a Schwarzschild space should be used,
but the Minkowski--Newton approximation is reasonably accurate away
from the centre and gives good
%%%qualitative    Rev1.2
results for the orbits.
Assume that the inertial frame at $P$ (at distance $r$ from the
origin) is rotating \wrt\ the background with angular velocity
$\omega(r)$ counter-clockwise.  When computing rotation curves, the
formula for $\omega(r)$ just found \eqref{eq:nett} will be used but
for the present discussion it is just as easy to assume a general
function.  The inertial frame at $P$ can be identified with the
background space, but it is important to remember that it is rotating.
There is no sensible meaning to the centre of rotation for an inertial
frame.  Two rotations which have the same angular velocity but
different centres differ by a uniform linear motion and inertial
frames are only defined up to uniform linear motion.  Thus it can be
assumed for simplicity that all the rotations have centre at the
origin.  Then the inertial frames can be pictured as layered
transparent sheets, each comprising the same point-set but with each
one rotating with a different angular velocity about the origin.  Each
sheet corresponds to a particuar value of $r$.  It is necessary to be
very clear about the nature of motion in one of these frames.  A
particle moving with a frame (ie one stationary in that frame) has no
\emph{inertial velocity} and its velocity is called \emph{rotational}.
In general if a particle has velocity $\mathbf{v}$ (measured in the
background space) then
$$
\mathbf{v}=\mathbf{v}_{\rm rot}+\mathbf{v}_{\rm inert}
$$ where its \emph{rotational velocity} $\mathbf{v}_{\rm rot}$ is the
velocity due to rotation of the local inertial frame and
$\mathbf{v}_{\rm inert}$ is its \emph{inertial velocity} which is the
same as its velocity measured \emph{in} the local inertial frame.
Note that $\mathbf{v}_{\rm rot} = r\omega(r)$ directed along the
tangent.

Inertial velocity correlates with the usual Newtonian concepts of
centrifugal force and conservation of angular momentum.

\sh{The fundamental relation}

As a particle moves in the equatorial plane it moves between the
sheets so that a rotation about the origin which is rotational in one
sheet becomes partly inertial in a nearby sheet.  For definiteness,
suppose that $\omega(r)$ is a decreasing function of $r$ and consider
a particle moving away from the origin and at the same time rotating
counter-clockwise about the origin.  The particle will appear to be
being rotated by the sheet that it is in and this causes a tangential
acceleration.  This acceleration is called the \emph{slingshot effect}
because of the analogy with the familiar effect of releasing an object
swinging on a string.  But at the same time the particle is moving to
a sheet where the rotation due to \id\ is decreased and hence part of
the tangential velocity becomes inertial and is affected by
conservation of angular momentum which tends to decrease the angular
velocity.  These two effects balance each other out in the limit and
this explains the flat asymptotic behaviour.  More precisely, let $v$
be the tangential velocity of the particle in the direction of
$\omega$ then the slingshot effect causes an acceleration $dv/dr =
\omega(r)$ but conservation of angular momentum that operates on the
inertial part of $v$, namely $v-r\thinm\omega(r)$ causes a deceleration in
$v$ of $(v-r\thinm\omega(r))/r$ or an acceleration $dv/dr = \omega(r) - v/r$
and adding the effects we have the \emph{fundamental relation} between
$v$ and $r$:
\begin{equation}\label{eq:fund}
\fboxrule1pt\fboxsep5pt\fbox{$\displaystyle\frac{dv}{dr}=2\omega-\frac vr$}
\end{equation}

Given $\omega$ as a function of $r$, \eqref{eq:fund} can be solved to
give $v$ as a function of $r$.  Rewrite it as
$$r\,\frac{dv}{dr}+v=2\omega r\,.$$
The LHS is $d/dr\,(rv)$ and the general solution is
\begin{equation}\label{eq:gensol}
v=\frac1r\Bigl(\int2\omega r\,dr + \text{const}\Bigr).
\end{equation}
It is now clear that any prescribed differentiable rotation curve can
be obtained by making a suitable choice of continuous $\omega$.

%%%% Rev1.3 new remark
\rk{Remark}It is worth remarking that the fundamental relation does
not depend on the {WSP}.  It is a purely geometric result depending only
on axial symmetry and the fact that the dynamic is controlled by a
central force.  It is true in any metrical theory and in particular in
general relativity without the {WSP}.  However for the models constructed here
it will be used with the formula for $\omega$ found in \eqref{eq:nett}
which does depend on the WSP.
\medskip

Of interest here are solutions which, like observed rotation curves,
are asymptotically constant, and, inspecting \eqref{eq:gensol}, this
happens precisely when $\int2\omega r\,dr$ is asymptotically equal to
$Qr$ for some constant $Q$.  But this happens precisely when $2\omega$
is asymptotically equal to $Q/r$.  This proves the following result.

\begin{theorem}The equatorial geodesics have tangential velocity
  asymptotically equal to constant $Q$ if and only if $\omega$ is
  asymptotically equal to $A/r$ where $Q=2A$.
\end{theorem}

\sh{The basic model}

Now specialise to the case $\omega=\lfrac{A}{(r+K)}$ which gives the
value of \id\ formulated in \eqref{eq:nett}.

\begin{figure}[ht!]
\labellist\small
\pinlabel 2 [r] at 14 457
\pinlabel 1 [r] at 14 247
\pinlabel 10 [t] at 130 17
\pinlabel 20 [t] at 252 17
\pinlabel 30 [t] at 364 17
\pinlabel 40 [t] at 463 17
\endlabellist
\cl{\includegraphics[width=.4\hsize]{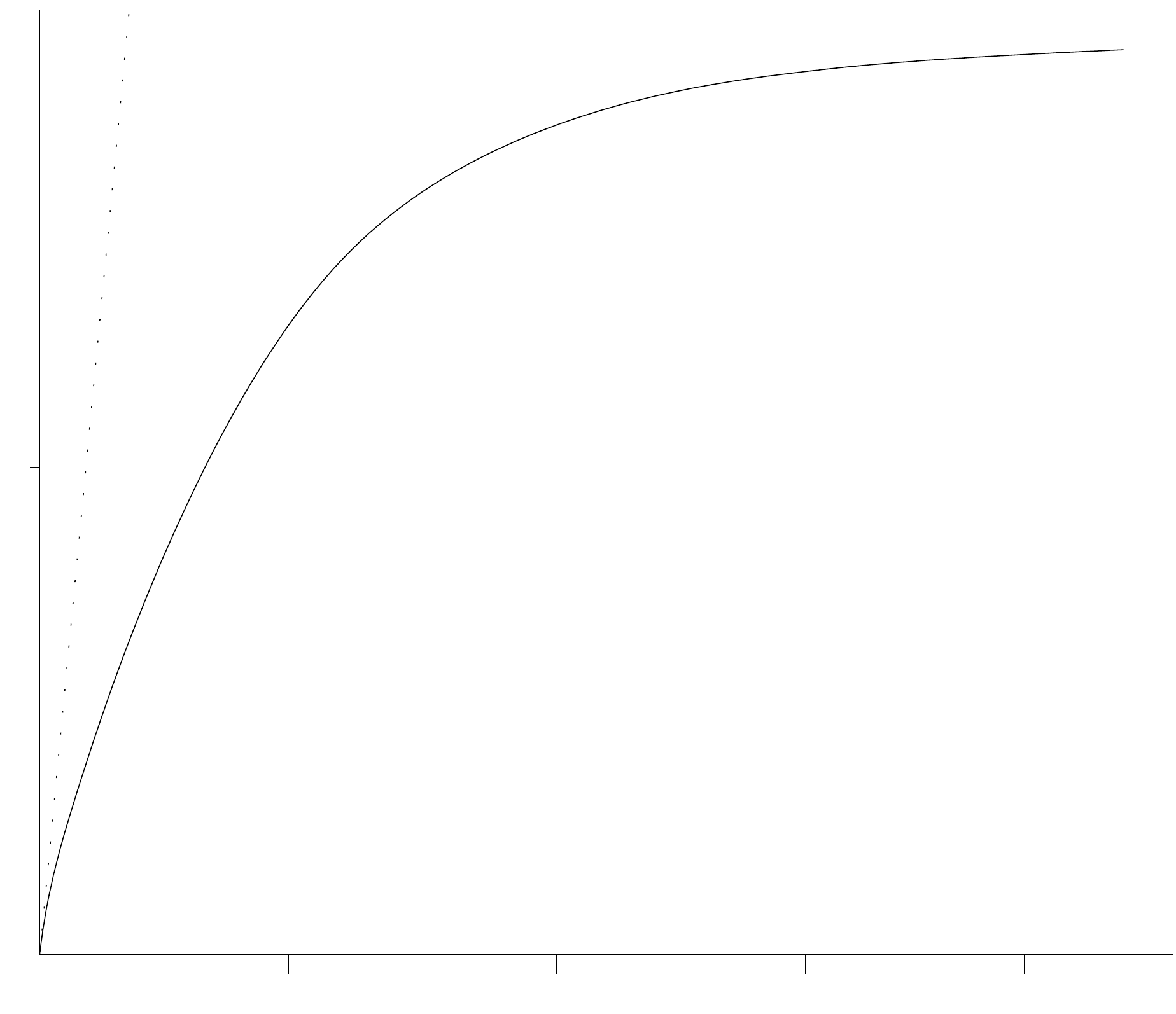}\qquad\includegraphics[width=.4\hsize]{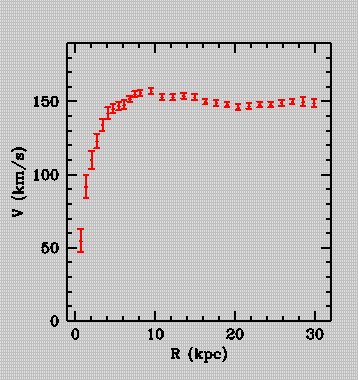}} 
\caption{The rotation curve from the model (left) and for the galaxy
  NGC3198 (right) taken from Begeman \cite{Beg}}
\label{fig:modbeg}
\end{figure}
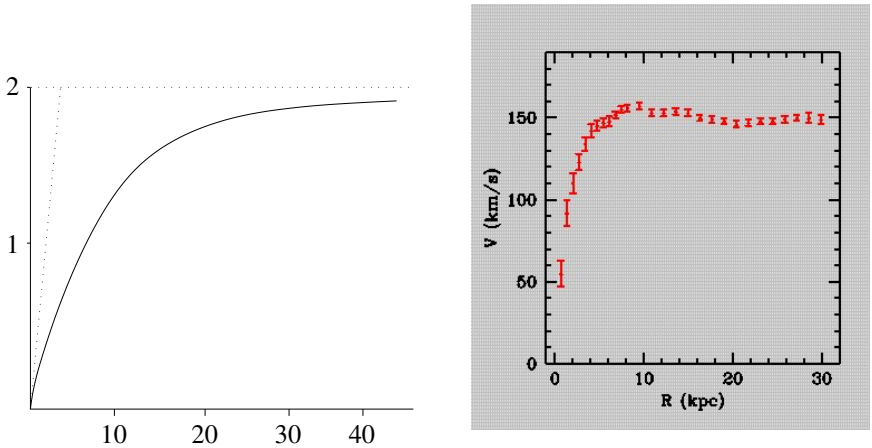

\begin{figure}[ht!]
\cl{\includegraphics[width=.7\hsize]{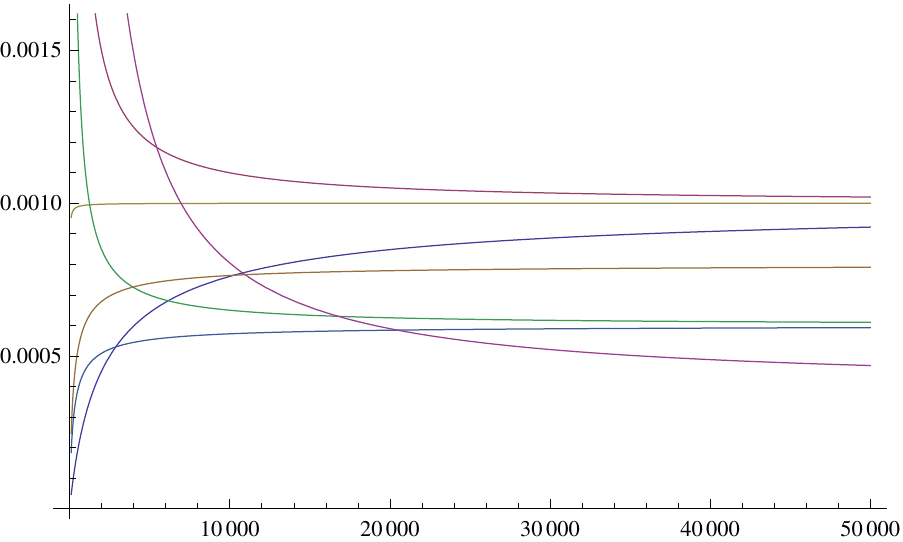}}
\caption{A selection of rotation curves from the model}
\label{fig:rotsmod}
\end{figure}

From \eqref{eq:gensol}
\def\strutt{\vrule width 0pt height 20pt}
\begin{align}
v&=\frac1r\left(\int\frac{2Ar}{r+K}\,dr +
C\right)=\frac{2A}r\left(\int1-\frac{K}{r+K}\,dr\right) + \frac Cr\notag\\
 &= 2A -\frac {2AK}r \log\left(\frac rK +1\right) + \frac Cr\strutt\label{eq:v}
\end{align}
where $C$ is a constant depending on initial conditions.  For a
particle ejected from the centre with $v= r\omega_0$ for $r$ small,
$C=0$, and for general initial conditions there is a contribution
$C/r$ to $v$ which does not affect the behaviour for large $r$.  For
the solution with $C=0$ there are two asymptotes.  For $r$ small,
$v\approx r\omega_0$ and the curve is roughly a straight line through
the origin.  And for $r$ large the curve approaches the horizontal
line $v=2A$.  A rough graph is given in \fullref{fig:modbeg} (left)
where $K=A=1$.  The similarity with a typical rotation curve,
\fullref{fig:modbeg} (right), is obvious.  Note that no attempt has
been made here to use meaningful units on the left.  See
\fullref{fig:rotsmod} below for curves from the model using sensible
units.

But notice that every equatorial orbit has the salient feature of
observed rotation curves, namely a horizontal asymptote.  This
asymptote is \emph{the same for all equatorial orbits} and hence any
average over many orbits will also have this asymptote and this
explains the observed rotation curve.

\sh{Units}  %%%% Rev 1.4: discussion of units

The models for the rotation curve and for galactic dynamics given in
this paper are fully \emph{quantitative} using \emph{natural} units,
in which $G$ (Newton's gravitational constant) and $c$ (the velocity
of light) are both $1$.  Time is measured in years, distances in
light-years, velocities in fractions of $c$ and mass converted into
distance using the Schwarzschild radius: a mass of 1 means a mass with
Schwarzschild radius 1 (light-year).  Thus a velocity of .001 is
$300\mathrm{km/s}$, a distance of 45,000 is 15Kpc and a mass of 1 is
$3\times10^{12}$ solar masses all approximately.  When using natural
units, pure numbers are used.  They can be converted into more
familiar units as indicated here.

\begin{figure}[ht!]
\cl{\includegraphics[height =1.8in]{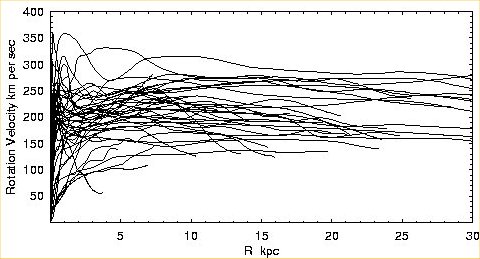}}
\caption{A collection of rotation curves  from \cite{SR}}
\label{fig:rotsSR}
\end{figure}

There are other shapes for rotation curves; see \cite{SR} for a
survey.  All agree on the characteristic horizontal straight line.
\fullref{fig:rotsSR} is reproduced from \cite{SR} and gives a good
selection of rotation curves superimposed.  In \fullref{fig:rotsmod}
is a selection of rotation curves again superimposed, sketched using
Mathematica\fnote{The notebook {\tt Rots.nb} used to draw this figure
  can be collected from \cite{Nb} and the values of the parameters
  used read off.} and the model given here.  The central masses for
the curves in \fullref{fig:rotsmod} vary from $3\times10^{11}$ to
$10^{14}$ solar masses.  The different curves correspond to choices of
$A,K$ and $C$.  The similarity is again striking.

It is worth commenting that the observed rotation curve for a galaxy
is not the same as the rotation curve for one particle, which is what
has been modelled here.  When observing a galaxy, many particles are
observed at once and what is seen is a rotation curve made from
several different rotation curves for particles, which may be close
but not identical.  So it is expected that the observed rotation
curves have variations from the modelled rotation curve for one
particle, which is exactly what is seen in \fullref{fig:modbeg}
(right) and \fullref{fig:rotsSR}. 

\sh{Postscript}

As remarked earlier, the effect described in this section is
independent of mass.  However for rotating bodies of small mass the
effect is unobservably small.  For example the sun has $K\approx\mathrm{3km}$ ,
assuming $K=M$, and $\omega_{\rm Sun}=2\pi/25$ days.  Thus the
asymptotic tangential velocity $2A=2K\omega_{\rm Sun}$ is $\mathrm{6km}$ per 4
days or $\mathrm{1m/s}$ approximately.

\section{The full dynamic}\label{sec:dyn}

The analysis given in \fullref{sec:rot} will now be extended to find
equations for orbits in general (not just for the tangential velocity)
and, using a hypothesised central generator, the spiral arm structure
will be modelled as well.  The basic idea is that the central mass
accretes a belt of matter which develops instability and explodes
feeding the roots of the arms.  Stars are formed by condension in the
arms and move outwards as they develop.  Thus a typical star is on a
long outward orbit and the rotation curve observed for stars in a
spiral arm is formed of many such similar orbits.  But this full
picture is not necessary to explain the observed rotation curves,
since the tangential velocity for all orbits has the same horizontal
asymptote.  

\rk{Note}The assumption of a hypermassive central black hole in a
spiral galaxy directly contradicts current beliefs of the nature of
Sagittarius \astar\ and this problem together with other observational
matters are dealt with in Chapter 6 of the book \cite{Book} of which
this paper is a fragment.  Very briefly, Sgr\astar\ and the stars in
close orbit around it form an old globular cluster near the end of its
life with most of the matter condensed into the central black hole.
It is not at the centre of the galaxy but merely roughly on line to
the centre and it is about half-way from the sun to the real galactic
centre which is invisible to us.

\sh{Plotting orbits}

Equation \eqref{eq:v} gave a formula for
the tangential velocity in an orbit.  What is needed is a formula for
the radial velocity (again in terms of $r$) and these two will
describe the full dynamic in the equatorial plane, which can then be
used to plot orbits.

\begin{figure}[ht!]
\cl{\includegraphics[width=.6\hsize]{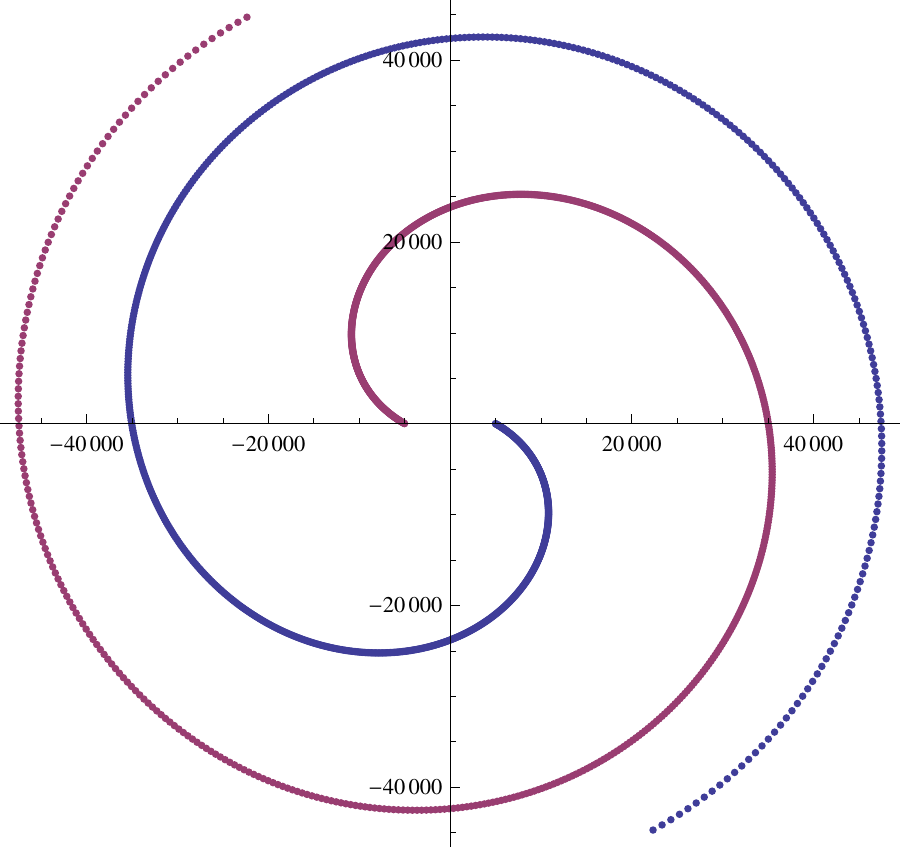}}
\caption{Mathematica plot using equations \eqref{eq:energy} and \eqref{eq:v-recap}}
\label{fig:basic}
\end{figure}

There are two radial ``forces'' on a particle: a centripetal force
because of the attraction of the massive centre and a centrifugal
force caused by rotation in excess of that due to inertial drag.  Thus
radial acceleration $\rdd$ is given by
\begin{equation}\label{eq:rdd}
\rdd = \frac{v_{\inert}^2}r - F(r)
\end{equation}
where $v_{\inert}=v-\omega r$ and $F(r)$ is the effective central
``force'' at radius $r$, per unit mass, which, since we are using a
Newtonian approximation, is $M/r^2$.  The same notation as in the last
section is used here and in particular $\omega=\omega(r)$ is the
\id\ at radius $r$.  Now specialise to the case
$\omega=\lfrac{A}{(r+K)}$, equation \eqref{eq:nett}, which was the
formula for \id\ coming from the Weak Sciama Principle.  Here
$A=K\omega_0$ and $K=kM$, where $M,\omega_0$ are the mass and angular
velocity of the central mass, and $k$ is a weighting constant which
can be taken to be 1 for purposes of exposition.  The following
formula for $v$, equation \eqref{eq:v}, was found: \def\strutt{\vrule
  width 0pt height 20pt}
\begin{equation}
v= 2A -\frac {2AK}r \log\left(\frac rK +1\right) + \frac Cr\strutt\label{eq:v-recap}
\end{equation}
where $C$ is a constant which can be read from the tangential velocity
for small $r$.  This implies:
\begin{equation}\label{eq:vinert}
v_{\inert}=2A -\frac {2AK}r \log\left(\frac rK +1\right) + \frac Cr\strutt - \frac{Ar}{K+r}
\end{equation}
Thus:
$$\rdd = \frac{v_{\inert}^2}r - \frac M{r^2} = \frac1r\Bigg[2A -\frac {2AK}r \log\left(\frac rK +1\right) + \frac Cr\strutt - \frac{Ar}{K+r}\Bigg]^2-\frac M{r^2}$$
Multiplying by $\rdot$ and integrating \wrt\ $t$ gives
\begin{align}\label{eq:energy}
\tfrac12 \rdot^2 = \int \rdd dr =& - \frac{C^2}{2r^2} + \frac{M - 2AC}r +
\frac{A^2K}{K + r} + A^2\log(K + r) \\
&+   \frac{2AK(C + 2Ar)\log(1 + r/K)- (2 AK\log(1 + r/K))^2}{r^2} + E\notag
\end{align}
where $E$ is another constant determined by the overall energy of the
orbit.  From this equation $\rdot$ can be read off (in terms of $r$).
Moreover since there is a formula for $v$, there is also a formula for
$\dot\theta = v/r$, where polar coordinates $(r,\theta)$ are used in
the equatorial plane.  From this it is possible to express $\theta$ and
$t$ in terms of $r$ as integrals.  These integrals are not easy to
express in terms of elementary functions but Mathematica can be used to
integrate them numerically and this can be used to plot the orbits of
particles ejected from the centre.  Now use the hypothesis that the
centre of a normal galaxy contains a belt structure, which emits jets
of gas$/$plasma, which condense into stars.  The orbits of these stars
can be modelled and a ``snapshot'' of all the orbits taken at an
instant of time, in other words a picture of the galaxy can be given,
\fullref{fig:basic}.

\begin{figure}[ht!]
\cl{\includegraphics[height=1.9in]{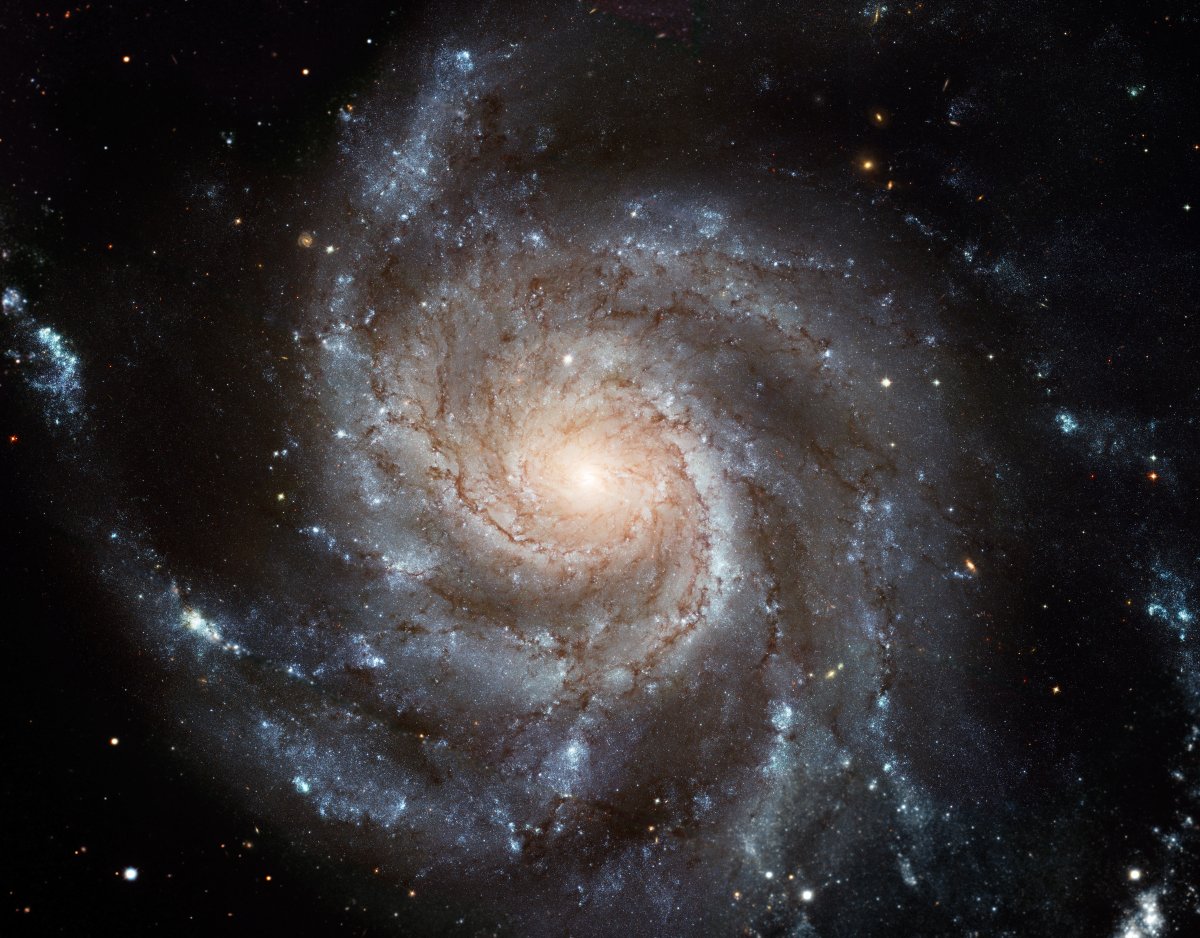}\quad\includegraphics[height=1.9in]{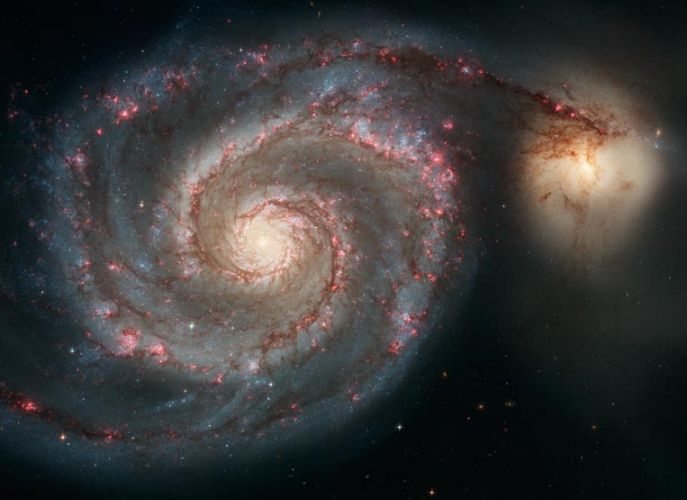}}
\caption{M101 (left) and  M51 (right): images from the Hubble site
\cite{Hubble}} \label{fig:classic-spirals}
\end{figure}

This compares well with classic observed spiral galaxies,
\fullref{fig:classic-spirals}. 

\iffalse
%%%% Rev 1.5 new end to section and new next section
In the next section we reproduce the Mathematica program used to plot 
\fullref{fig:basic} and explain how to use it.  The program can be
downloaded from \cite{Nb} as \texttt{Basic.nb} and the reader is
encouraged to experiment with it.
\fi
To end this section it is worth remarking that it may be thought that
motion along the arms is a new hypothesis and that observations should
be able to verify this.  However \textit{the observations aleady
  exist}.  The rotation curve shows that there is a general outward
motion and, since the arms are close to tangential, as seen in the
above figures, this motion is roughly along the arms, which is
therefore already observed.  The new hypothesis is that this motion is
responsible for the long-term maintenance of the spiral structure.
The model demonstrates all of this both graphically and
quantitatively.

\section{A new paradigm for the universe}
As noted earlier, this paper is an extract from the author's book ``A
new paradigm for the universe'' \cite{Book}, which covers all the
topics discussed here in greater detail.  The justification for the
key factor $1/r$ in the main geometric hypothesisis is given in detail
in \cite[Chapter 2]{Book}.  The derivation of the fundamental relation
\eqref{eq:fund} and the radial acceleration equation \eqref{eq:rdd}
are both proved in detail for a larger class of suitable metrics in
\cite[Chapters 3 and 5]{Book}.  The proposed accretion structure (the
``generator'') which creates the jets which condense into the spiral
arms is described in more detail in \cite[Chapter 5]{Book}.  The
Mathematica notebook used for the plot, \fullref{fig:basic}, is typed
out (and can be downloaded from \cite{Nb} as \texttt{Basic.nb}) and
the data used spelt out as follows.

$2A$ the asymptotic tangential velocity is $.001$ which is 300km/s in
MKS units.  $M$ has been set to $10^{11}$ solar masses.  The minimum
radius for the plot, {\tt rmin}, has been set to 5,000 and the
maximum, {\tt rmax}, to 50,000 light years (corresponding to a visible
diameter of 100,000 light years).  There is a precession constant $B$
which arises because inertial drag causes the frame at the origin to
appear to rotate at $A/${\tt rmin} and $B=A/${\tt rmin} corrects this
and causes the familiar spiral form.  Time elapsed along the visible
arms is $5.5\times 10^7$ years.  The nature of the visible spiral arms
is discussed carefully in \cite[Chapter 6]{Book}.  Here merely note
that the visible arms correspond to strong star-producing regions and
bright short life stars, which burn out or explode in $10^5$ to $10^7$
years.  Thus a total time elapsed of $5.5\times 10^7$ years allows
several generations of stars to be formed and to create the heavy
elements necessary for planets such as the earth to be formed.

\rk{Remark}The rate of rotation of the central mass $\omega_0$ can be
read from the other variables.  Since $A=K\omega_0$, from equation
\eqref{eq:nett}, and the default value $K=M$ has been chosen,
$\omega_0 = A/M = .0005/.03 = 1/60$.  In other words the central mass
is rotating at about one radian every 60 years or one revolution every
360 year approx.  This ought to be observable, but since the central
mass in a galaxy is obscured by the bulge, all that can be observed
are the velocities of stellar regions near the centre and these are
dominated by the rotation curve.  Since the correct rotation curve is
built into the model, observations here are in agreement with the
model.
\medskip

The program is intended for interactive use and the reader is
recommended to download a copy and investigate the output.  Hints on
using it can be found in \cite[Section 5.6]{Book}.

The book also covers a wide range of other topics and in particular
gives a new framework for quasars, which fits with the observations of
Halton Arp \cite{Arp} and others, that quasars typically exhibit
intrinsic redshift, and explains the apparently paradoxical results of
Hawkins \cite{H}.

\end{document}